\documentclass[prl,twocolumn,superscriptaddress,nofootinbib]{revtex4}
\usepackage{graphicx}
\usepackage{dcolumn}
\usepackage{multirow}
\usepackage{amsmath,amsthm,amssymb}
\usepackage{subfigure}
\usepackage{url}
\usepackage{hyperref}

\DeclareMathAlphabet{\mathpzc}{OT1}{pzc}{m}{it}

\begin{document}
\title{Resonant dynamics and the instability of anti-de Sitter spacetime}

\author{Piotr Bizo\'n}
\affiliation{Institute of Physics, Jagiellonian University, Krak\'ow, Poland}
\author{Maciej Maliborski}
\affiliation{Max-Planck-Institut f\"ur Gravitationsphysik,
Albert-Einstein-Institut, Golm, Germany}
\author{Andrzej Rostworowski}
\affiliation{Institute of Physics, Jagiellonian University, Krak\'ow, Poland}
\date{\today}
\begin{abstract}
  We consider spherically symmetric Einstein-massless-scalar field equations with negative cosmological constant in five dimensions and analyze  evolution of small perturbations of anti-de Sitter spacetime using the recently proposed resonant approximation. We show that for typical initial data the solution of the resonant system develops an oscillatory singularity in finite time. This result hints at a possible route
  to establishing  instability of AdS under arbitrarily small perturbations.

\end{abstract}

\maketitle
 \textit{Introduction}.
 A few years ago two of us gave numerical evidence that anti-de Sitter (AdS) spacetime in four dimensions is unstable against black hole formation for a large class of arbitrarily small perturbations \cite{br}. More precisely, we showed that
for a perturbation  with amplitude $\varepsilon$ a black hole forms on the timescale $\mathcal{O}(\varepsilon^{-2})$. Using nonlinear perturbation analysis we conjectured  that the instability is due to the turbulent cascade of energy from low to high frequencies. This conjecture was extended to higher dimensions in \cite{jrb}.

  Since the computational cost of numerical simulations rapidly increases with decreasing $\varepsilon$,   our conjecture was based on extrapolation of the observed scaling behavior of solutions for small (but not excessively so) amplitudes, which left some room for doubts whether the instability will persist to arbitrarily small values of $\varepsilon$ (see e.g. \cite{dfly}).  To resolve these doubts, in this paper we validate and reinforce the above extrapolation with the help of a recently proposed resonant approximation \cite{lehner, cev1, cev2}. The key feature of this approximation is that the underlying infinite dynamical system (hereafter referred to as the resonant system) is scale invariant:  if its solution with amplitude $1$ does something at time $t$, then the corresponding solution with amplitude $\varepsilon$ does the same thing at time $t/\varepsilon^2$. Moreover, the latter solution remains close to the true solution (starting with the same initial data) for times $\lesssim \varepsilon^{-2}$ (provided that the errors due to omission of higher order terms do not pile up too rapidly).
 Thus, by solving the resonant system we can probe the regime of arbitrarily small perturbations (whose outcome of evolution is beyond the possibility of numerical verification).

  For concreteness, in this paper we focus our attention on AdS$_5$ (the most interesting case from the viewpoint of  AdS/CFT correspondence);  an extension to other dimensions is straightforward and will be presented elsewhere.
\vskip 0.1cm
 \textit{Model}. For the reader's convenience, let us recall from \cite{br,jrb} the general framework for studying the spherically symmetric scalar perturbations of the AdS spacetime.
 The five dimensional  asymptotically AdS spacetimes are parametrized by the coordinates $(t,x,\omega)\in (-\infty,\infty) \times [0,\pi/2) \times \mathbb{S}^3$ and the metric
\begin{equation}
\label{ads5}
ds^2\! =\! \frac {\ell^2}{\cos^2{\!x}}\left( -A e^{-2 \delta} dt^2 + A^{-1} dx^2 + \sin^2{\!x} \, d\omega^2\right)\,,
\end{equation}
where $\ell^2=-6/\Lambda$, $d\omega^2$ is the round metric on $\mathbb{S}^3$,  and $A$, $\delta$ are functions of $(t,x)$.
 For this ansatz the evolution of a self-gravitating massless scalar field $\phi(t,x)$ is governed by the following system (using units in which $8\pi G=3$)
\begin{align}
\label{eq_wave}
\partial_t \Phi & = \partial_x \left( A e^{-\delta} \Pi \right), \quad \partial_t \Pi = \frac{1}{\tan^{3}{\!x}} \partial_x \left(\tan^{3}{\!x} \,A e^{-\delta} \Phi \right),\\
\label{eq_A}
\partial_x A \!&= \!\frac{2+2\sin^2{\!x}} {\sin{x}\cos{x}} \, (1-A) - \sin{x}\cos{x} \, A \left( \Phi^2 + \Pi^2 \right),
\\
\label{eq_delta}
\partial_x \delta \!&=\! -  \sin{x}\cos{x} \left( \Phi^2 + \Pi^2 \right),
\end{align}
where
$\Phi= \partial_x \phi$ and $\Pi= A^{-1} e^{\delta} \partial_t \phi$.
To ensure smoothness at spatial infinity and finiteness of the total mass $M$ we impose the boundary conditions  (using $\rho=\pi/2-x$)
\begin{align}\label{pi2}
    \phi(t,x)&= f_{\infty}(t)\, \rho^4+\mathcal{O}\left(\rho^{6}\right),\quad
    \delta(t,x)=\delta_{\infty}(t)+ \mathcal{O}\left(\rho^{8}\right), \nonumber \\
    A(t,x)&=1- M \rho^4+\mathcal{O}\left(\rho^{6}\right),
\end{align}
where the power series expansions are uniquely determined by $M$  and  the functions $f_{\infty}(t)$, $\delta_{\infty}(t)$. We use the normalization $\delta(t,0)=0$, hence $t$ is the proper time at the center.
We will solve this system for small smooth perturbations of the  AdS solution $\phi=0, A=1, \delta=0$.
\vskip 0.1cm
 \textit{Resonant approximation}.
  As follows from  equation \eqref{eq_wave}, linearized perturbations of AdS$_5$  are governed by the  operator $L=-\tan^{-3}{\!x}\, \partial_x (\tan^3{\!x} \, \partial_x)$. This operator is essentially self-adjoint with respect to the inner product  $(f,g):=\int_0^{\pi/2} f(x) g(x) \tan^3{\!x}\, dx$. The eigenvalues and orthonormal eigenfunctions of $L$  are $\omega_n^2=(2n+4)^2$ ($n=0,1,\dots$) and
  \begin{equation}\label{eigenfun}
  e_n(x)=2\sqrt{\frac{(n+2)(n+3)}{n+1}} \, \cos^4{\!x}\, P_n^{(1,2)}\left(\cos(2x)\right)\,,
  \end{equation}
  where $P_n^{(1,2)}(x)$ is the Jacobi polynomial of order $n$.

After these preliminaries we are prepared to introduce the resonant approximation. To avoid technicalities, let us first illustrate this approach in the case of the cubic wave equation on the \emph{fixed} AdS$_5$ background\footnote{The study of this equation  has been argued (see \cite{basu} and references therein)  to provide  insight into the problem of stability of AdS. We do not share this view (as explained in \cite{b}) and  use equation \eqref{u3}  solely for illustrative purposes, \emph{not} as a toy model.}
\begin{equation}\label{u3}
  \partial_{tt} \phi +L \phi +\sec^2{\!x}\, \phi^3 =0\,.
\end{equation}
Inserting  the mode expansion
$\phi(t,x)=\sum_n c_n(t) e_n(x)$ into \eqref{u3} we get an infinite system of coupled oscillators
\begin{equation}\label{fourier}
  \frac{d^2 c_n}{dt^2} +\omega_n^2 c_n = \sum\limits_{jkl} I_{jkln}\, c_j c_k c_l\,,
\end{equation}
where  the coefficients $I_{jkln}=-(e_j e_k e_l \sec^2{\!x},e_n)$ determine interactions between the modes.
To factor out fast linear oscillations in \eqref{fourier}, we change variables using the variation of constants (a.k.a.``interaction picture")
\begin{eqnarray}\label{voc}
  c_n &=&\beta_n e^{i\omega_n t} +\bar \beta_n e^{-i\omega_n t},\\
  \frac{d c_n}{dt} &=& i\omega_n \left(\beta_n e^{i\omega_n t} -\bar \beta_n e^{-i\omega_n t}\right)\,.
\end{eqnarray}
This transforms the system \eqref{fourier} into
\begin{equation}\label{eqs}
 2 i \omega_n \frac{d \beta_n}{dt} =  \sum\limits_{jkl} I_{jkln} \,
 c_j c_k c_l \, e^{-i\omega_n t}\,,
 \end{equation}
 where  each $c_j$ in the sum is given by (9), thus each term in the  sum
  has a factor $e^{-i\Omega t}$, where $\Omega=\omega_n\pm \omega_j\pm \omega_k \pm \omega_l$. \\The terms with $\Omega=0$ correspond to resonant interactions, while those with $\Omega\neq 0$ are non-resonant.

  Passing to slow time $\tau=\varepsilon^2 t$ and  rescaling $\beta_n(t)=\varepsilon  \alpha_n(\tau)$, we see that
for  $\varepsilon$ going to zero the non-resonant terms   $\propto e^{-i\Omega \tau/\varepsilon^2}$ are highly oscillatory and therefore negligible (at least for some  time). Keeping only the resonant terms in \eqref{eqs} (which  is equivalent to time-averaging),  we obtain  the infinite autonomous dynamical system (which we shall call the \emph{resonant system})
\begin{equation}\label{rs-u3}
  2 i \omega_n \frac{d \alpha_n}{d\tau} = \sum\limits_{j k l} I_{jkln} \,\alpha_j \alpha_k \bar \alpha_l \,,
\end{equation}
where the summation  runs over the set of indices $\{j,k,l\}$ for which $\Omega=0$ and $I_{jkln}\neq 0$ (due to the fully resonant non-dispersive spectrum of $L$ and vanishing of some coefficients $I_{jkln}$, this set reduces to $\{jkl\, |\, j+k-l=n\}$; see footnote 3 in \cite{cev2}).  Note that the system \eqref{rs-u3} is invariant  under the scaling $\alpha_n(\tau) \rightarrow \varepsilon^{-1} \alpha_n(\tau/\varepsilon^2)$.
  It is routine to show that the solutions of \eqref{eqs} starting from small initial data of size $\varepsilon$ are well approximated  by the solutions of \eqref{rs-u3} on the timescale $\mathcal{O}(\varepsilon^{-2})$\footnote{More precisely, if $\beta_n(t)$ and $\alpha_n(\tau)$ are solutions of \eqref{eqs} and \eqref{rs-u3}, respectively, and $\beta_n(0)=\varepsilon \alpha_n(0)$ for each $n$, then $|\beta_n(t)-\varepsilon \alpha_n(\varepsilon^2 t)| \lesssim \varepsilon$ for $t \lesssim \varepsilon^{-2}$.}.
  In other words, on this timescale the dynamics of solutions of the cubic wave equation \eqref{u3} is dominated by resonant interactions.

  For the system (2-4) the derivation of the resonant system is similar but technically more intricate because, to begin with, one has to integrate out the constraints which (at the lowest order) results in nonlocal cubic nonlinearities in the derivatives of $\phi$. Despite these complications, the resonant system has the same form as \eqref{rs-u3}, namely
  \begin{equation}\label{rs}
  2 i \omega_n \frac{d \alpha_n}{d\tau} = \sum\limits_{j+k-l=n} C_{jkln} \,\alpha_j \alpha_k \bar \alpha_l \,,
\end{equation}
   except that now the interaction coefficients $C_{jkln}$ are  given by much more complicated  expressions involving integrals of products of eigenfunctions \eqref{eigenfun} and their derivatives.  The system \eqref{rs} was first derived  in \cite{lehner} and \cite{cev1} using the multiscale perturbation methods\footnote{In the multiscale approach,  the system \eqref{rs} follows from elimination of secular terms due to resonances at the third order of  perturbation expansion. The fact that all secular terms can be removed in this way has been sometimes misunderstood as evidence for stability of AdS.} 
   and soon afterwards in \cite{cev2} using the averaging method.
 It is simpler than the full system, yet still too difficult to be handled by purely analytic means, hence in what follows we analyze it using numerical and asymptotic methods.

\vskip 0.1cm
\textit{Results}.
We  solved in parallel the full Einstein equations (2-4) and the resonant system \eqref{rs} for a variety of the same small initial data. To illustrate the results (which we believe are universal), we present them for the two-mode initial data with
energy (almost) equally distributed among the modes
\begin{equation}
  \label{2-mode}
  \phi(0,x) = \varepsilon \, \left(\frac{1}{4} e_{0}(x)
    + \frac{1}{6} e_{1}(x)\right),\quad
  \Pi(0,x) = 0\,.
\end{equation}
Let us point out that,  since  small one-mode data and their perturbations  enjoy  quasiperiodic evolutions \cite{br, mr, lehner}, the two-mode data are in a sense minimal in what it takes to initiate the turbulent cascade (cf.\cite{br3}).

The numerical simulations of the full Einstein equations (2-4) were previously reported in \cite{jrb} in the case of Gaussian initial data. The evolution of the two-mode data \eqref{2-mode} looks similar. For all considered small values of $\varepsilon$ we observe collapse to a black hole in time $t_{H}(\varepsilon) \sim \varepsilon^{-2}$ (see Fig.~1).  This scaling suggests that the instability should  be seen  in the resonant approximation. In what follows,  we confirm this expectation and thereby give support to the conjecture that the instability is present for arbitrarily small perturbations.

\begin{figure}[!h]
  \centering
  \includegraphics[width=\columnwidth]
  {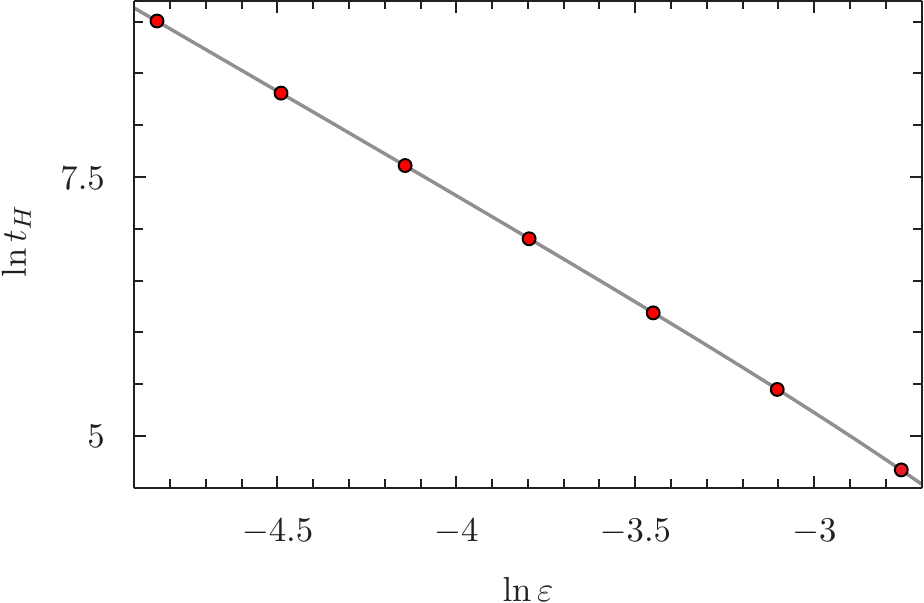}
  \caption{Time of horizon formation vs  amplitude in the evolution of initial data \eqref{2-mode}.
  The solid line depicts the fit of the function $-2\ln{\varepsilon}+a+b \varepsilon^2$ to the numerical data $\ln{t_H(\varepsilon)}$.
  From this fit we obtain $\tau_H:=\lim_{\varepsilon \rightarrow 0} \varepsilon^{-2} t_H(\varepsilon)=e^a \approx 0.514$.}
  \label{fig1}
\end{figure}

  For the numerical computation, the resonant system~\eqref{rs} must be truncated at some (possibly large) index~$N$.
 As a compromise between the accuracy\footnote{See \cite{krol} for the analysis of the combined error due to truncation and averaging for semilinear wave equations.} and the computational cost, we choose here $N=172$. To solve the truncated resonant system (TRS) numerically we use the 6th-order Gauss-Runge-Kutta method\footnote{Due to the oscillatory character of solutions,  explicit schemes turn
out to be unstable. More details of the numerical method and its validation are given in the supplementary material accompanying this note.}.
 The initial data for the TRS corresponding to \eqref{2-mode} are
\begin{equation}
  \label{rs-id}
  \alpha_{0}(0) = 1/8,
  \quad \alpha_{1}(0) = 1/12,
  \quad (\alpha_{n}(0))_{n\geq 2} = 0.
\end{equation}
 To describe and analyze the behavior of solutions, it is convenient to use the amplitude-phase representation $\alpha_n=A_n e^{i B_n}$ in terms of which the resonant system \eqref{rs} takes the following form \cite{cev2}
\begin{align}\label{eqA}
 & 2 \omega_n \frac{dA_n}{d\tau}  =  \sum\limits_{\substack{j+k-l=n\\j\neq n, k\neq n}} S_{jkln} A_j A_k A_l \sin(B_n+B_l-B_j-B_k) \\
&  2 \omega_n \frac{dB_n}{d\tau}  =  T_n A_n^2 +\sum\limits_{j\neq n} R_{jn} A_j^2 \nonumber \\
 & +  A_n^{-1} \sum\limits_{\substack{j+k-l=n\\j\neq n, k\neq n}} S_{jkln} A_j A_k A_l \cos(B_n+B_l-B_j-B_k)\,.
\end{align}
where $T_n=C_{nnnn}$, $R_{jn}=C_{njjn}$ if $j\neq n$, and $S_{jkln}=C_{jkln}$ if all four indices are different. Explicit expressions for these coefficients are given in the appendix A in~\cite{cev2}.

\begin{figure}[!h]
  \centering
  \includegraphics[width=\columnwidth]
  {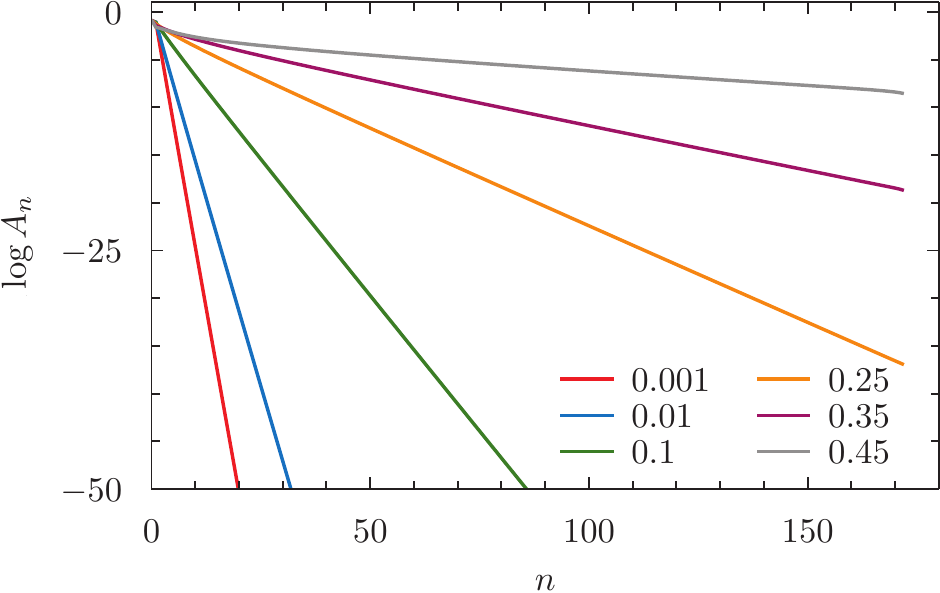}
  \caption{The  amplitude spectra (for several times $\tau$).}
  \label{fig2}
\end{figure}
\begin{figure}[!h]
  \centering
  \includegraphics[width=\columnwidth]
  {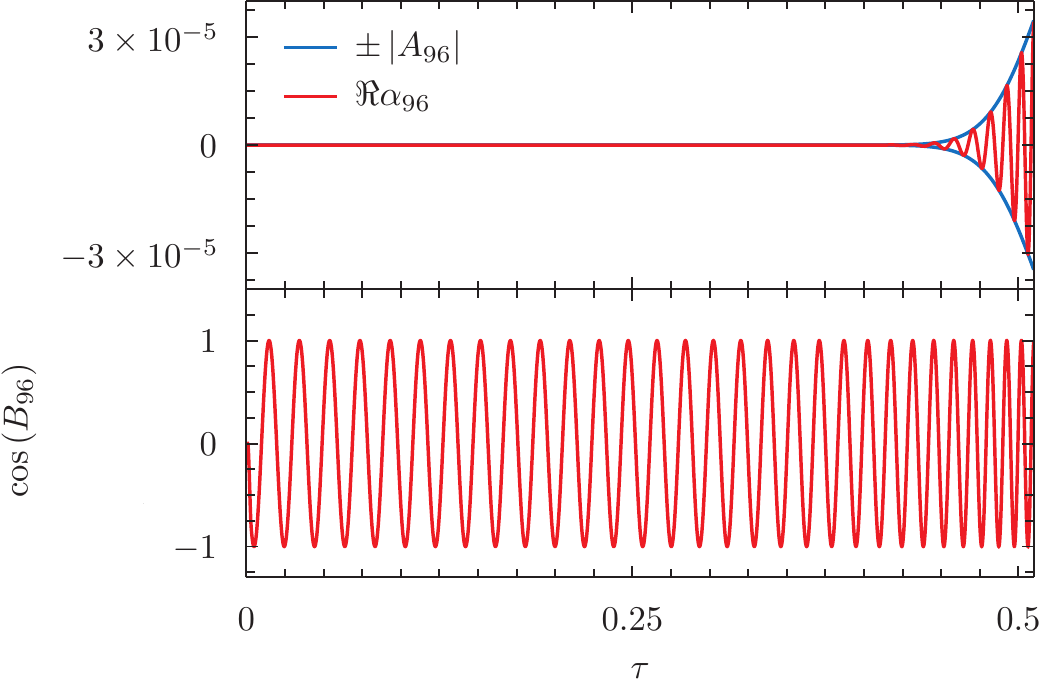}
  \caption{Evolution of a sample high mode.}
  \label{fig3}
\end{figure}
Evolving the data \eqref{rs-id}, we find that higher modes are quickly excited (see Fig.~2).
For early times the amplitudes grow at the polynomial rate $A_n(\tau)\sim\tau^{n-1}$ while the phases evolve approximately linearly. At a  later time the frequencies  of oscillations begin to grow rapidly (see Fig.~3).  This highly oscillatory behavior accumulates at a finite time causing numerical difficulties. We find that the time-step  of numerical integration, for which the algorithm is convergent,   tends to zero as the cutoff $N$ increases. This suggests that the solution of the resonant system ($N=\infty$) develops an oscillatory singularity in some finite time $\tau_*$\footnote{For any finite $N$  the solution of TRS cannot blow up and, with sufficient resolution, can be numerically continued  past $\tau_*$, however this `afterlife' is an artifact of truncation.}.

To give better evidence for blowup and analyze its character we proceed in the spirit of the analyticity strip method \cite{ssf, bj}, namely we make the following asymptotic ansatz for the amplitudes
\begin{equation}\label{ansatz}
  A_n(\tau) \sim  n^{-\gamma(\tau)} e^{-\rho(\tau) n} \quad \mbox{for}\,\,\, n\gg 1\,.
\end{equation}
Fitting this formula to the numerical data  we obtain the time dependence of the exponent $\gamma(\tau)$ and the `analyticity radius' $\rho(\tau)$. As shown in Fig.~4, it appears that $\rho(\tau)$ tends to zero in a finite time $\tau_*$ (with $\rho_0=-\rho'(\tau_*)>0$) confirming   that the solution of  the resonant system \eqref{rs} becomes singular at $\tau_*$. The fit  also reveals that the asymptotic power-law amplitude spectrum has the exponent $\lim_{\tau \rightarrow \tau_*} \gamma(\tau)=2$.

\begin{figure}[!h]
  \centering
  \includegraphics[width=\columnwidth]
  {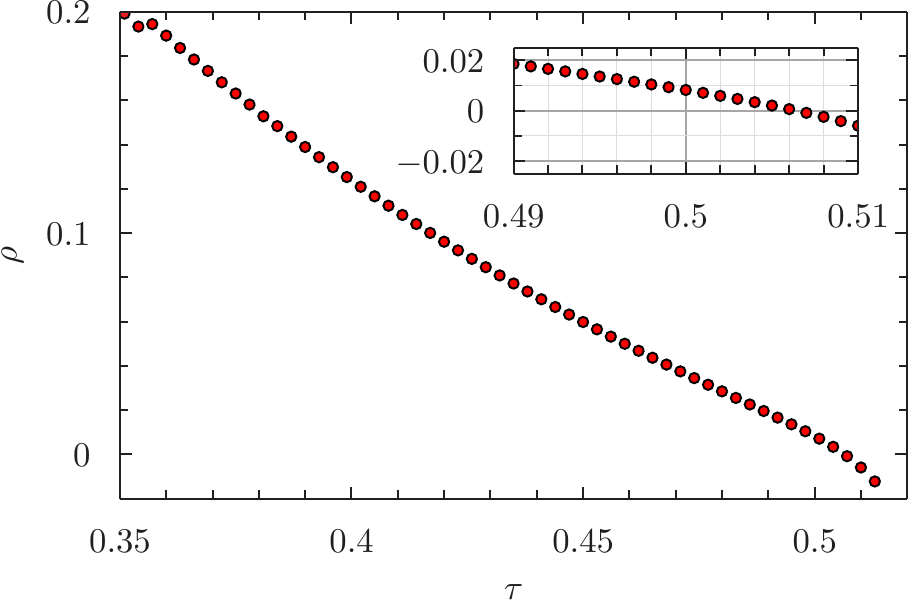}
  \caption{The radius of analyticity $\rho(\tau)$ obtained by fitting the formula (\ref{ansatz}) to the amplitude spectrum. The point of this (notoriously poor) fit is to show that $\rho(\tau)$ hits zero in some finite time $\tau_*$, not to determine $\tau_*$ precisely (cf. footnote 6).}
  \label{fig4}
\end{figure}

Guided by these numerical findings we will now construct an asymptotic solution of the resonant system that becomes singular in finite time. We assume that  for large $n$ and  $\tau \rightarrow \tau_*$
\begin{equation}\label{An}
  A_n(\tau) \sim n^{-2} e^{-\rho_0 (\tau_*-\tau) n}\,.
\end{equation}
To solve for  the phases, we note  the following asymptotic behavior of the interaction coefficients
\begin{equation}\label{c-asym}
  T_n \sim n^5,\quad R_{jn} \sim n^2 j^3, \quad S_{\lambda j,\lambda k, \lambda l, \lambda n} \sim \lambda^4\, S_{jkln}\,.
\end{equation}
Notice that the latter implies that
\begin{equation} \label{cancel}
\sum\limits_{\substack{j+k-l=n\\j\neq n, k\neq n}} S_{jkln} (j k l)^{-2} = \mathcal{O}(1)\,,
\end{equation}
that is the sum does not dependent of $n$.
Plugging \eqref{An} into Eq.(17) and using \eqref{c-asym} and \eqref{cancel}, we see that
 for $\tau \rightarrow \tau_*$ the r.h.s. of Eq.(17) is dominated by the term
\begin{equation}\label{sumR}
  \sum R_{jn} A_j^2 \sim n^2 \sum j^{-1} e^{-2\rho_0 (\tau_*-\tau) j} \sim n^2 \ln(\tau_*-\tau)\,,
\end{equation}
 thus the derivatives $\frac{dB_n}{d\tau}$ blow up logarithmically. Moreover, it follows from the above that for large $n$ the phases $B_n$ behave linearly with $n$, hence
$B_n+B_l-B_j-B_k \approx0$ for the resonant quartets. This implies that  both sides of Eq.(16) are (approximately) independent of $n$, reassuring that the ansatz \eqref{An} is self-consistent. Numerical simulations indicate that the asymptotics of blowup just described is in fact universal; this is illustrated  in Fig.~5 in the case of two-mode data \eqref{rs-id}.
\begin{figure}
  \centering
  \includegraphics[width=\columnwidth]
  {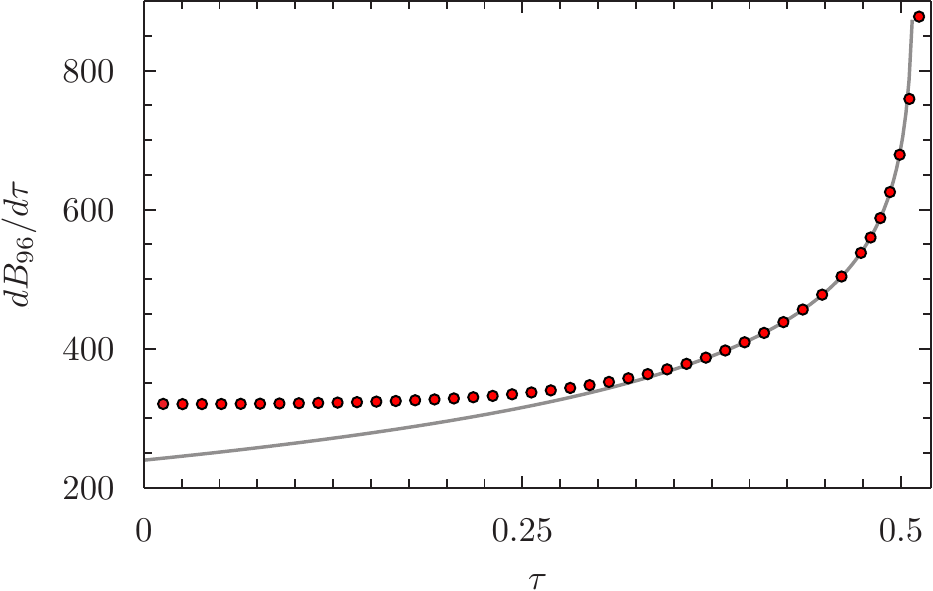}
  \caption{Evidence for the logarithmic blowup. The solid line represents the fit of the theoretical prediction $a_n \ln(\tau_*-\tau) +b_n$ to the numerical data $\frac{dB_n}{d\tau}$ for $n=96$. Performing this fit for  all  $n>20$ we find, in accord with the asymptotic analysis,  that the coefficients $a_n$ and $b_n$ vary linearly with $n$, while $\tau_*\approx 0.509$ does not depend on $n$. Note that $\tau_*$ agrees very well with the time of collapse $\tau_H$ given in the caption of Fig.~1.}
  \label{fig5}
\end{figure}

\textit{Conclusion}. To summarize, we have constructed the asymptotic solution of the resonant system that becomes singular in finite time and gave numerical evidence that this solution acts as a universal attractor for blowup. The key question is how to transfer this blowup result from the resonant system to the full system. On one hand, we see that the resonant approximation reproduces the amplitudes of true solutions remarkably  well  almost all the way to collapse. This is  illustrated in Figs.~6 and 7 where we compare the energy spectrum and the growth of the Ricci scalar at the origin computed  in parallel using the full Einstein equations  and the resonant system. On the other hand, the resonant approximation does not
work so well for the phases\footnote{Note that the highly oscillatory behavior of solutions is in tension  with the idea of time averaging.}. For this reason (and because of possible breakdown of the cubic approximation near collapse), it is not clear to us what (if any)  is the physical interpretation of the oscillatory singularity for the resonant system\footnote{Since the resonant system has no scale, it is tempting to speculate that there is a relationship between the oscillatory singularity and Choptuik's discretely self-similar solution \cite{ch}.}. Nonetheless, the fact  that solutions of the resonant system blow up in finite time (for typical initial data) strongly indicates that the corresponding solutions of the full system collapse on the timescale $\mathcal{O}(\varepsilon^{-2})$.

\begin{figure}
  \centering
  \includegraphics[width=\columnwidth]
  {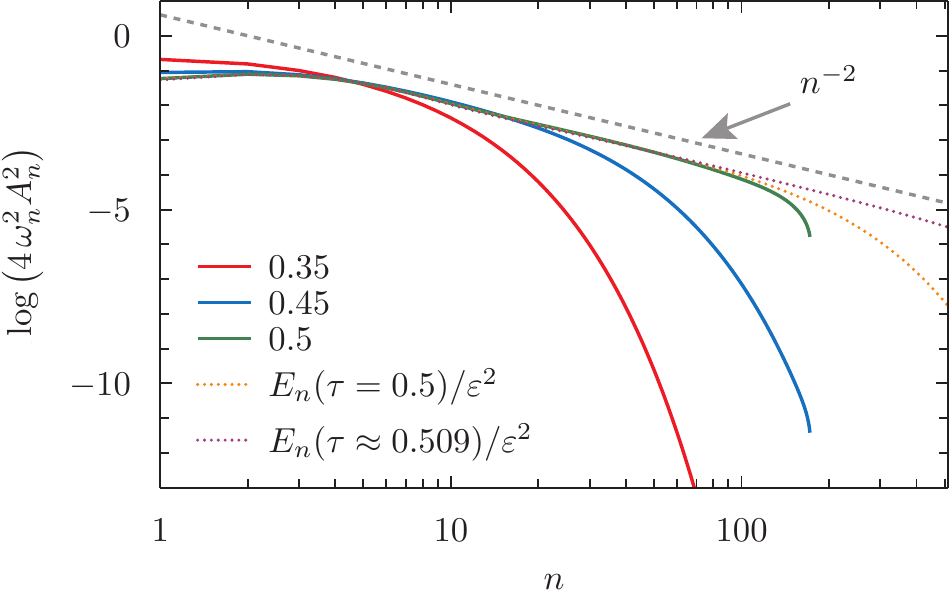}
  \caption{Energy spectra for the full system (dotted lines) and TRS (solid lines) at sample late times (cf. Fig.~2). The power-law spectrum $n^{-2}$ unfolds as the solutions collapse/blow up. Here $\varepsilon \approx 0.0079$ (the smallest amplitude used in simulations).}
  \label{fig6}
\end{figure}
\begin{figure}
  \centering
  \includegraphics[width=\columnwidth]
  {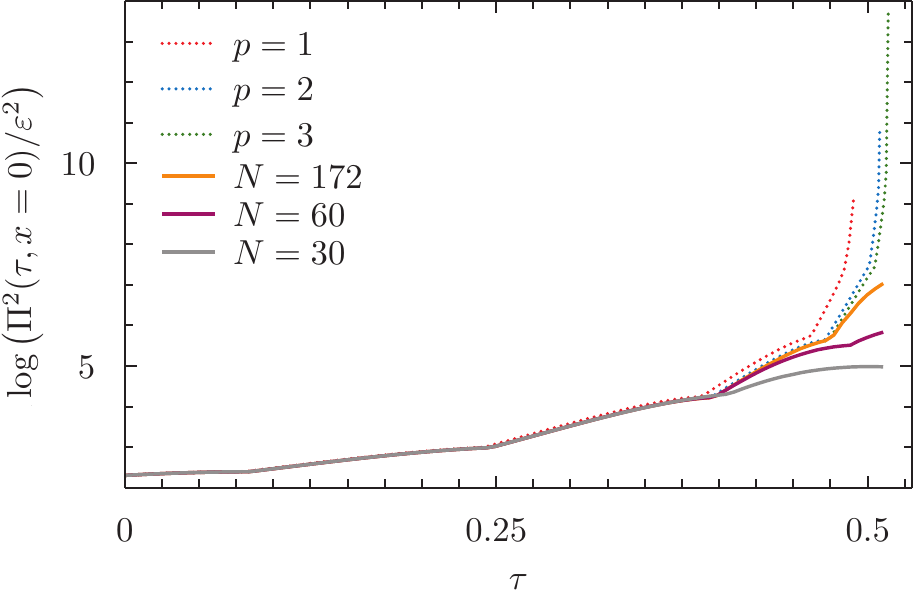}
  \caption{Upper envelope of  $\Pi^2(t,0)$  in the evolution of  initial data \eqref{2-mode} with $\varepsilon=(2\pi)^{-3/2} 2^{-p}$ for $p=1,2,3$ (dotted lines).  As $\varepsilon$ decreases, the rescaled quantities
    $\varepsilon^{-2} \Pi^{2}(\varepsilon^2 t,0)$ approach  a  limiting curve. The corresponding solutions of  TRS (solid lines) appear to approach the same curve as $N$ increases.}
  \label{fig:EnergySpectra2}
\end{figure}

\noindent \emph{Acknowledgments:} This work was supported  in part by the Polish National Science Centre Grant no. DEC-2012/06/A/ST2/00397. The computations were carried out with the supercomputer ``Zeus'' maintained by Academic Computer Centre CYFRONET AGH (Grant no. MNiSW/Zeus\_lokalnie/UJ/027/2014).


\begin{thebibliography}{10}

\bibitem{br}
P. Bizo\'n, A. Rostworowski,
Phys. Rev. Lett. \textbf{107}, 031102 (2011), \href{http://arxiv.org/abs/1104.3702}{\texttt{[arXiv:1104.3702]}}


\bibitem{jrb}
J. Ja\l{}mu\.zna, A. Rostworowski, P. Bizo\'n,
Phys. Rev. D \textbf{84}, 085021 (2011), \href{http://arxiv.org/abs/1108.4539}{\texttt{[arXiv:1108.4539]}}


\bibitem{dfly}
F.V. Dimitrakopoulos, B. Freivogel, M. Lippert, I.-S. Yang,
%
\href{http://arxiv.org/abs/1410.1880}{\texttt{[arXiv:1410.1880]}}


\bibitem{lehner}
V. Balasubramanian, A. Buchel, S.R. Green, L. Lehner, S.L. Liebling,
%
Phys. Rev. Lett. \textbf{113}, 071601 (2014), \href{http://arxiv.org/abs/1403.6471}{\texttt{[arXiv:1403.6471]}}

\bibitem{cev1}
B. Craps, O. Evnin, J. Vanhoof,
%
JHEP \textbf{10}, 48 (2014),
\href{http://arxiv.org/abs/1407.6273}{\texttt{[arXiv:1407.6273]}}


\bibitem{cev2}
B. Craps, O. Evnin, J. Vanhoof,
%
JHEP \textbf{01}, 108 (2015), \href{http://arxiv.org/abs/1412.3249}{\texttt{[arXiv:1412.3249]}}



\bibitem{basu} P. Basu, C. Krishnan, P.N. Bala Subramanian, \href{http://arxiv.org/abs/1501.07499}{\texttt{[arXiv:1501.07499]}}

\bibitem{b} \url{http://www.ctc.cam.ac.uk/activities/adsgrav2014/Slides/Slides_Bizon.pdf}

\bibitem{mr} M. Maliborski, A. Rostworowski, Phys. Rev. Lett.  111, 051102 (2013), \href{http://arxiv.org/abs/1403.5434}{\texttt{[arXiv:1403.5434]}}

\bibitem{br3} P. Bizo\'n, A. Rostworowski, \href{http://arxiv.org/abs/1403.5434}{\texttt{[arXiv:1410.2631]}}

\bibitem{krol} M.S. Krol, Math. Methods Appl. Sciences 11, 649 (1989)

\bibitem{ssf} C. Sulem, P.-L. Sulem, H. Frisch, J. Comput. Phys. 50, 138 (1983)

\bibitem{bj} P. Bizo\'n, J. Ja\l mu\.zna,
Phys. Rev. Lett. 111, 041102 (2013), \href{http://arxiv.org/abs/1306.0317}{\texttt{[arXiv:1306.0317]}}

\bibitem{ch} M.W. Choptuik, Phys. Rev. Lett. 70, 9 (1993)

\end{thebibliography}
\end{document}